\documentclass[letterpaper]{article} 
\usepackage{aaai25}  
\usepackage{times}  
\usepackage{helvet}  
\usepackage{courier}  
\usepackage[hyphens]{url}  
\usepackage{graphicx} 
\usepackage{natbib}  
\usepackage{caption} 
\usepackage{algorithm}
\usepackage{algorithmic}

\usepackage{newfloat}
\usepackage{listings}
\DeclareCaptionStyle{ruled}{labelfont=normalfont,labelsep=colon,strut=off} 
\floatstyle{ruled}
\newfloat{listing}{tb}{lst}{}
\floatname{listing}{Listing}
\usepackage[utf8]{inputenc}
\usepackage{graphicx}
\usepackage{mathrsfs}
\usepackage{amsmath}
\usepackage{amsthm}
\usepackage{soul}
\usepackage{bm}
\usepackage{booktabs}
\usepackage[switch]{lineno}
\usepackage{multirow}
\usepackage{wasysym}
\usepackage{color}

\usepackage{subfig}
\usepackage{amsfonts}

\newcommand{\longname}{\textit{\underline{G}reen \underline{R}ecommender \underline{A}ligned with \underline{P}ersonalized \underline{E}ating}}
\newcommand{\shortname}{\textit{GRAPE}}

\title{Bites of Tomorrow: Personalized Recommendations for a Healthier and Greener Plate}
\author{
    Jiazheng Jing$^*$,
    Yinan Zhang\thanks{These two authors contributed equally to this work.},
    Chunyan Miao\thanks{Corresponding author.}
}

\affiliations{
    College of Computing and Data Science (CCDS), Nanyang Technological University (NTU), Singapore\\
    jiazheng001@e.ntu.edu.sg, \{yinan.zhang, ascymiao\}@ntu.edu.sg
}

\usepackage{bibentry}

\begin{document}

\maketitle

\begin{abstract}

The recent emergence of extreme climate events has significantly raised awareness about sustainable living. In addition to developing energy-saving materials and technologies, existing research mainly relies on traditional methods that encourage behavioral shifts towards sustainability, which can be overly demanding or only passively engaging. In this work, we propose to employ recommendation systems to actively nudge users toward more sustainable choices. We introduce \longname{} (\shortname{}), which is designed to prioritize and recommend sustainable food options that align with users' evolving preferences. We also design two innovative \textit{Green Loss} functions that cater to green indicators with either uniform or differentiated priorities, thereby enhancing adaptability across a range of scenarios. Extensive experiments on a real-world dataset demonstrate the effectiveness of our \shortname{}.
\end{abstract}

%

\section{Introduction}

The recent emergence of extreme climate events, including the record-breaking heatwaves in July 2024, catastrophic floods, and increasingly frequent wildfires, indicates a troubling shift toward global instability and highlights the critical urgency of addressing climate change \cite{jain2022observed, ngcamu2023climate}. Significant research efforts have been conducted to develop proactive solutions towards sustainability, including the development of renewable technologies and materials that enhance energy efficiency \cite{samir2022recent, fang2024carbon}. Additionally, innovations in water purification, desalination, and recycling technologies are being explored to ensure a sustainable water supply\cite{darvishi2023comprehensive, curto2021review}.

Since human activities play an important role in global sustainability, some research works concentrate on understanding and encouraging behavioral shifts towards sustainability~\cite{kirby2021sustainability, marcus2019search}. For example, \cite{Lenka2023} explore the influence of factors such as environmental attitudes and lifestyle on online purchasing behavior, identifying significant impacts from an environmentally oriented lifestyle and willingness to pay for green products. Regarding behavioral nudges, most existing works focus on implementing regulations and designing poster campaigns to promote sustainable practices \cite{peleg2022regulation}. These methods, while traditional, can sometimes be perceived as imposing or may rely on passive engagement, which might not consistently engage or motivate the target audience.

Recommendation systems have shown considerable potential to deliver effective, personalized, and timely nudges that can lead to meaningful shifts in user behavior across various domains \cite{he2024impact}. Given the vast amount of products and services available, it is often impractical for consumers to review the whole item pool. Instead, they typically make choices from a curated list of options presented to them. By learning from users' historical behaviors and ongoing feedback, recommendation systems can prioritize sustainable items that align with user preferences, thereby effectively guiding consumers towards more sustainable choices \cite{zhang2024greenrec}. However, most existing works simply aim to improve prediction accuracy or enrich user experience, often neglecting the potential of recommendation systems to influence user behavior towards sustainability~\cite{zhang2024multi, lin2024multi}.

In this work, we introduce a novel task for food recommendation systems, named \textit{Green Food Recommendation}: to recommend foods that not only appeal to users but also actively prioritize and promote more sustainable options. We focus on the food domain for several reasons. First, food production is associated with multiple environmental issues, including water pollution, carbon emissions, and the use of arable land \cite{clark2022estimating}. Second, food is a fundamental part of daily life, and even minor sustainable improvements can lead to significant impacts \cite{poore2018reducing}. Third, food sustainability encompasses a variety of criteria, including environmental impact, nutritional value, and more. By integrating multiple sustainability indicators, the recommendation approaches becomes more robust and versatile, applicable across various domains.

Several challenges need to be addressed in \textit{Green Food Recommendation}. First, individual attitudes towards different aspects of sustainability can vary and change over time. Simply increasing the exposure to more sustainable food options may lead to user dissatisfaction. Therefore, it is crucial to recommend items that balance both evolving personal preference and sustainability. Moreover, the multiple indicators associated with food sustainability complicate the optimization of recommendations. These indicators, which reflect various aspects of food, can vary significantly. For example, although fish is nutritious, it often scores poorly on environmental impact indicators due to the water pollution caused during its cultivation.

To tackle the aforementioned challenges, we propose a novel green food recommendation method named \longname{} (\shortname{}). To model users' varying preferences and attitude towards sustainability, we employ both self-attention and cross-attention mechanisms. To encourage more sustainable food choices, we designed two novel \textit{Green Loss} functions that prioritizes more sustainable options. Regarding the \textit{Green Loss} functions, one treats all sustainability indicators equally, while the other assigns them different priorities. Extensive experiments on a real-world food dataset \cite{zhang2024greenrec} were conducted to evaluate the proposed \shortname{}. Experimental results demonstrate that \shortname{} outperforms all state-of-the-art baselines in making both accurate and sustainable recommendations.

With this paper, we make the following contributions:
\begin{itemize}
    \item We propose a novel research task of \textit{Green Food Recommendation} to encourage further research that not only caters to user preferences but also nudges user behaviors towards sustainability. 
    \item To bridge the gap between users' preferences and item sustainability, we propose \longname{} (\shortname{}). In \shortname{}, we designed two innovative \textit{Green Loss} functions that address green indicators with either equal or differentiated priorities, enhancing adaptability in various scenarios.
    \item Extensive experimental results highlight the superior performance of \shortname{} in making accurate and sustainable recommendations. Furthermore, our findings demonstrate that it is feasible to maintain recommendation accuracy while enhancing the sustainability attributes of the recommended food. We believe that our research establishes a solid groundwork for future studies in this area.

\end{itemize}

\section{Related Work}
Recent years have witnessed the worsening of environmental issues such as global warming and rising sea levels, which has brought increasing attention to sustainability issues from academia.\cite{jain2022observed, ruggerio2021sustainability}. 
Many studies focus on investigating individual environmental awareness, for example, to explore the environmental awareness between people from different backgrounds \cite{kirby2021sustainability}, and investigate the causes and future trends of people's sustainability awareness \cite{marcus2019search}. Besides, there are also researches focusing on exploring the current status and influence of specific environmental issues, for instance, climate change \cite{orlinska2021biomass}, mountain fire \cite{yin2024temporal}, and so on.
Additionally, solutions to environmental problems are also a key focus of research across various fields, including environmentally friendly materials \cite{samir2022recent} and industrial emission reduction \cite{fang2024carbon}.
These studies have made significant contributions to a sustainable society. However, through exploring the importance of individual behavior in addressing environmental issues, these studies have rarely delved deeply into the possibilities of encouraging people to live greener lifestyles.

Artificial intelligence (AI) has developed rapidly in recent years, showing great potential to solve problems in various fields, including leverage AI to solve sustainability problems, called Green AI \cite{schwartz2020green, verdecchia2023systematic}. 
For example,  semantic, segmentation and multimodal methods are utilized for satellite image processing \cite{To2024Carbon} and climate forecast \cite{jones2017machine}, conserving resources while also improving task accuracy. 
These efforts provide efficient solutions to specific problems. However, few works consider leveraging AI's influence on people's behavior to address sustainability issues.

Recommendation systems (RS), which are widely applied in modern society, have great potential to influence individual behaviors, which means that recommendation systems could be able to encourage and nudge users to adopt a greener lifestyle \cite{he2024impact, forouzandeh2024uifrs, rostami2023towards}. Notice this, Zhang et al. \cite{zhang2024greenrec} introduced a green dataset for food recommendations where each food is labeled with three sustainability indicators. Based on this dataset, CLUSSL \cite{zhang2024multi} is proposed to improve the recommendation performances utilizing sustainability indicators.
However, most of the work focuses only on improving recommendation accuracy, with very few efforts dedicated to providing greener recommendations.

\begin{figure*}[ht]
    \centering
	\includegraphics[width=2.1\columnwidth, trim=9 0 10 0, clip]{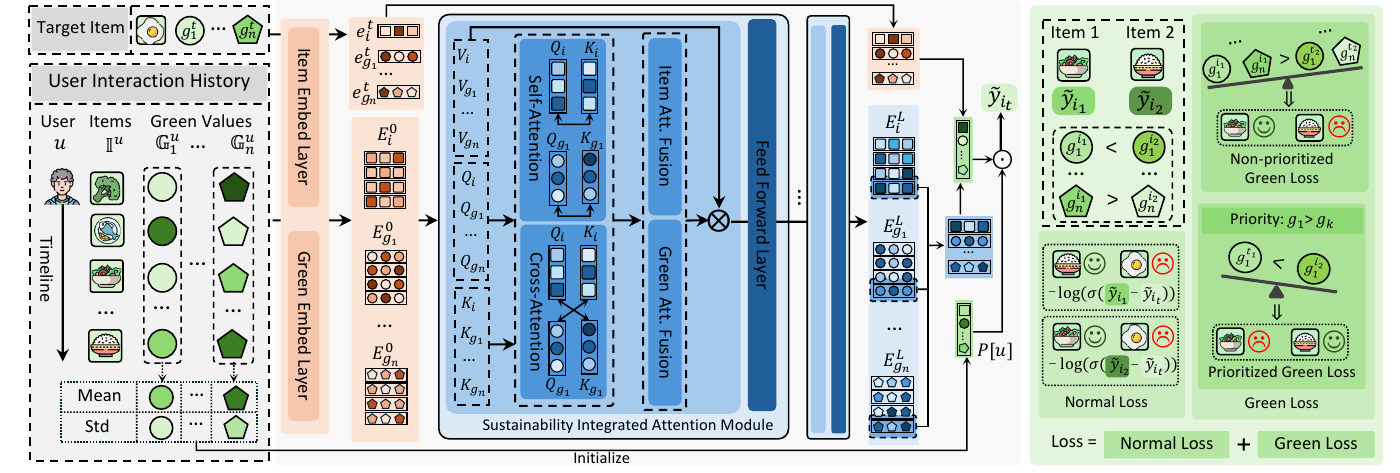}
	\caption{The model structure of \shortname{}. \shortname{} consists of three main modules: the Embedding Module, the Sustainability Integrated Attention Module, and the Prediction Module. The loss function is divided into two parts: the \textit{Normal Loss}, which aims to enhance recommendation accuracy, and the \textit{Green Loss}, which targets the improvement of recommendation sustainability. For the item's sustainability indicators, a deeper color signifies a greener value.}

	\label{fig:architecture}
\end{figure*}

\section{Method}

\subsection{\textit{Green Food Recommendation} Task}

Given an item $i \in \mathcal{I}$, there are $n$ sustainability indicators associated with $i$, labeled as $g_1^i, g_2^i, ..., g_n^i$, respectively. 
Given user set $\mathcal{U}$ with size $|\mathcal{U}|=v$, for user $u \in \mathcal{U}$ who have interacted with $w$ items, we denote the interacted item sequence as $\mathbb{I}^u=[i_1, i_2, ..., i_w]$, where $i_j$ represents the $j$-th interacted item in the chronologically ordered.
Similarly, the sequence of the corresponding sustainability indicators for $\mathbb{I}^u$ is denoted as $\mathbb{G}^u_1=[g_1^{i_1}, g_1^{i_2}, ..., g_1^{i_w}], ..., \mathbb{G}^u_n=[g_n^{i_1}, g_n^{i_2}, ..., g_n^{i_w}]$. 
The goal of the \textit{Green Food Recommendation} is to recommend items that not only satisfy the user's interests but also have high sustainability indicators. 


\subsection{Model Architecture}
Figure \ref{fig:architecture} shows the overall structure of our proposed \shortname{}, which consists of: Embedding Module, Sustainability Integrated Attention Module, and Prediction Module.

\subsubsection{Embedding Module} 
\label{Sec:Embedding_M}
In this module, we introduce two distinct embedding functions for the interacted items and their corresponding sustainability indicators.
First, we embed the sequence of items $\mathbb{I}^u$ that user $u$ has interacted with, using a one-hot embedding layer $\mathbf{L}_I$ to obtain embeddings ${E}_i^0 \in \mathbb{R}^{w\times d}$. Here, $d$ represents the embedding size.

For the sustainability sequences $\mathbb{G}^u_1, ...,\mathbb{G}^u_n$, we introduce a joint embedding layer $\text{JE}(\cdot)$ to discretize and embed the sustainability indicator $g_j^{i_k}$:
\begin{equation}
    \text{JE}(g_j^{i_k})=\lfloor \frac{g_j^{i_k}}{\Delta} \rfloor \times \mathbf{L}_G, j \in \{1, ..., n\}, k\in\{1,...,w\}, 
\end{equation}
  where $\mathbf{L}_G\in\mathbb{R}^{\frac{max(g_j)}{\Delta} \times d}$, $\Delta$ is a hyperparameter that controls the level of granularity in the discretization process, $\lfloor \cdot \rfloor$ denotes the floor function \cite{chen2022time}. We then construct the embedding matrix ${E}_{g_1}^0, ...,{E}_{g_n}^0$ by concatenating the corresponding indicator embeddings. 
\subsubsection{Sustainability Integrated Attention Module}
Inspired by \cite{lin2024multi}, we design a Sustainability Integrated Attention (SIA) Module to model the evolving preferences of users towards items and various sustainability indicators. Specifically, the SIA Module comprises four layers: a Projection Layer, Self- and Cross-Attention Layers, Attention Fusion Layers, and a Feed-Forward Layer. In \shortname{}, we stack $L$ SIA modules, and here we provide a detailed introduction to the $l$-th module.


\textbf{Projection Layer: }We first project each embedding matrix to capture query, key and value matrices with $M$ attention heads. The $m$-th attention head for query matrix projection is formulated as:
\begin{equation}
    Q^m_i = f_{Q^m_i}(E^{l-1}_i), Q^m_{g_j} = f_{Q^m_{g_j}}(E^{l-1}_{g_j}), j \in \{1, ..., n\},
\end{equation}
where $f_{Q^m_i}(\cdot), f_{Q^m_{g_j}}(\cdot)$ denote a linear projection layer, $g_j$ denotes the $j$-th sustainability indicator. Similarly, we have key matrices $K^m_i, K^m_{g_j}$ and value matrices $V^m_i, V^m_{g_j}$.

\textbf{Self- and Cross-Attention Layers: } Given that users' preferences change over time and their behaviors demonstrate sequential dependencies, capturing these sequential behaviors is crucial for making accurate predictions \cite{jing2023capturing,chen2022time}. In \shortname{}, we employ a self-attention mechanism to adeptly model users' evolving preferences for items and various sustainability indicators:
\begin{equation}
    A^m_x = Q^m_x{K^m_x}^\top, x\in\{i, g_1, ..., g_n\}.
\end{equation}

Furthermore, we observe strong correlations among the items users interact with and their corresponding different sustainability indicators \cite{zhang2024greenrec}. For instance, users who frequently purchase nutritional food also tend to prefer items with high Healthy Meal Index scores. To capture such correlations within users' interaction sequences, we introduce a cross-attention layer:
\begin{equation}
    A^m_{x,y} = Q^m_{x,y}{K^m_{x,y}}^\top, {x,y}\in\{i, g_1, ..., g_n\}, x\neq y.
\end{equation}

\textbf{Attention Fusion layers: }
We then incorporate fusion layers to model user preferences towards items and various sustainability indicators.
As for the item preference, we first aggregate all captured self- and cross-attention matrices:
\begin{equation}
    \mathcal{A}_i = \{A_x, A_{y,z}|x,y,z\in\{i, g_1, ...,g_n\},y\neq z\}.
\end{equation} 
Given $\mathcal{A}_i$, the Item Attention Fusion layer is formulated as
\begin{equation}
    R^m_{i} = \text{Mask}(\text{Fuse}(\bm{r}_{i}, \mathcal{A}_i), V^m_i)),
\end{equation}
where $\bm{r}_{i} \in \mathbb{R}^{{(n+1)}^2}$ is a learnable weight vector, $\text{Fuse}(\cdot)$ is a attention fusion function consists of weighted sum and gating operations \cite{liu2021noninvasive}, $\text{Mask}(\cdot)$ is a masked integration function which merges the captured attention matrix with value matrix $V^m_{i}$ without information leakage \cite{kang2018self}.

Regarding user's sustainable preferences, we combine all attention matrices that include all sustainability-related information:
\begin{equation}
    \mathcal{A}_g = \{A_x, A_{y,z}|x,y,z\in\{g_1, ...,g_n\},y\neq z\}.
\end{equation} 
Then the embedding for each sustainability indicator preference is generated through the Green Attention Fusion layer:
\begin{equation}
    R^m_{g_j} = \text{Mask}(\text{Fuse}(\bm{r}_{g}, \mathcal{A}_g), V^m_{g_j})), j \in \{1, ..., n\},
\end{equation}
where $\bm{r}_{g} \in \mathbb{R}^{{n}^2}$ is a learnable weight vector.



\textbf{Feed-Forward Layer: }
Finally, we leverage a Feed-Forward Network (FFN) to merge the output of all $M$ attention heads~\cite{vaswani2017attention}.
\begin{equation}
    \begin{array}{c}
       E^l_i=\text{FFN}([R^1_i, ..., R^M_i]W_i), \\
        E^l_{g_j}=\text{FFN}([R^1_{g_j}, ..., R^M_{g_j}]W_{g_j}), j \in \{1, ..., n\},
    \end{array}
\end{equation}
where $W_i, W_{g_j} \in \mathbb{R}^{d\times d}$ are learnable matrices.

We leverage $L$ SIA modules. For the $l$-th SIA module, the matrices ${E}^l_i, {E}^l_{g_j} (j \in \{1, ..., n\})$ are then defined as:
  \begin{equation}
      E_i^{l}, E_{g_1}^{l}, ..., E_{g_n}^{l} =\text{SIA}_l(E_i^{l-1}, E_{g_1}^{l-1}, ..., E_{g_n}^{l-1}),
  \end{equation}
where $\text{SIA}_l, l\in\{1, ..., L\}$ denotes the $l$-th SIA module.
We then extract the last embedding vector from each matrix as the final preference representations, denoted as ${\bm{e}}^L_i, {\bm{e}}^L_{g_j}$, respectively \cite{devlin2018bert}.

\subsubsection{Prediction Module}\label{Sec:Prediction_M}
Different users have varying preferences and levels of acceptance for sustainable food. Some might prefer junk food but also open to trying healthy options occasionally while some others may strictly favor junk food and show significant resistance to changing their dietary habits. In this module, we employ an attention layer to explicitly model users' preferences for items and their inclination to select items with high sustainability values.
Specifically, we first multiply each output of SIA modules $\bm{e}_i^L, \bm{e}_{g_1}^L,..., \bm{e}_{g_n}^L$ with the embeddings $\bm{e}_i^0, \bm{e}_{g_1}^0,..., \bm{e}_{g_n}^0$ obtained by Embedding Module, respectively. Then, we introduce an attention matrix $P\in\mathbb{R}^{v\times(n+1)}$:
\begin{equation}
    \widetilde{y}_ = [\bm{e}^L_i \bm{e}_i^{0\top}, \bm{e}^L_{g_1} \bm{e}_{g_1}^{0\top}, ...,\bm{e}^L_{g_n}\bm{e}_{g_n}^{0\top}]^\top\times P[u],
    \label{equ:predict}
\end{equation}
where $P[u] \in \mathbb{R}^{1\times (n+1)}$ is the $u$-th vector in $P$, representing $u$'s preference for the item itself and various sustainability dimensions.


\subsection{Training}
\subsubsection{Initialization}
We observe that the average values of sustainability indicators for items users have previously interacted with can somewhat reflect their preferences for sustainable food~\cite{zhang2024multi}. 
A higher average usually indicates that the user prefer more sustainable food.
Similarly, the standard deviation values of these indicators shed light on their acceptance. When users' daily eating habits are both relatively unhealthy and consistent, a larger variance may suggest a greater likelihood of users adopting more sustainable options.
To make full use of the users' interaction history and improve training, we propose to initialize the attention matrix $P$ using:
\begin{equation}
    P[u]=[1, \mu(\mathbb{G}^u_1)\lambda(\mathbb{G}^u_1), ..., \mu(\mathbb{G}^u_n)\cdot\lambda(\mathbb{G}^u_n)],
    \label{equ:init_P}
\end{equation}
where $\mu(\cdot)$ and $\lambda(\cdot)$ represent the functions for calculating the mean and variance of a sustainability indicator sequence, respectively.
$P[u]$ is regularize to meet $\sum P[u]=1$.

\subsection{Sampling and Loss Functions}
Given a user $u$, we use $\mathcal{I}_u^+, \mathcal{I}_u^-$ to denote the sets of items that user $u$ has interacted with and not interacted with, respectively.
The loss $\mathcal{L}$ contains two components: \textit{Normal Loss} $\mathcal{L}_n$ which helps \shortname{} generate accurate recommendations based on the user's interaction history, and \textit{Green Loss} $\mathcal{L}_g$, designed to enhance the visibility of more sustainable food options:
\begin{equation}
    \mathcal{L} =\alpha\mathcal{L}_n + (1-\alpha)\mathcal{L}_g,
    \label{Equa: all_loss}
\end{equation}
where $\alpha \in [0, 1]$ is a hyperparameter.

As for the \textit{Normal Loss} $\mathcal{L}_n$, we sample $i^+ \in \mathcal{I}_u^+$ and $i^- \in \mathcal{I}_u^-$ and leverage the pair wise BPR Loss \cite{rendle2012bpr}:
\begin{equation}
    \mathcal{L}_n = -\sum_{< u, i^+, i^- >} log(\sigma(\widetilde{y}_{i^+}- \widetilde{y}_{i^-})),
\end{equation}
where $\widetilde{y}_{i}$ denotes the predicted probability that user $u$ will interact with item $i$ as determined by \shortname{}, $\sigma$ refers to the sigmoid function.

The \textit{Green Loss} $\mathcal{L}_g$ is designed to assign higher scores to more sustainable items. However, it's crucial to align these scores with user preferences, rather than merely increasing the visibility of greener products. A recommendation is considered unsuccessful if a user dislikes the recommended item, regardless of its sustainability. Therefore, we differentiate between items the user has interacted with, $\mathcal{I}_u^+$, and those they haven't, $\mathcal{I}_u^-$, to ensure that our recommendations accurately reflect both sustainability and user preferences.
Specifically, we sample $i_1, i_2 \in \mathcal{I}_u^+$ or $i_1, i_2 \in \mathcal{I}_u^-$, where $[g_1^{i_1}, ..., g_n^{i_1}]$ represents the sustainability indicator values for $i_1$, and $[g_1^{i_2}, ..., g_n^{i_2}]$ represents those for $i_2$. We assume that a higher value of $g_j^i$ indicates that item $i$ is more sustainable according to the $j$-th sustainability indicator. For sustainability indicators where a smaller value signifies greater sustainability, we apply inverse conditions when calculating the loss.

We introduce two types of \textit{Green Loss} functions: the \textit{Non-prioritized Green Loss} $\mathcal{L}_{ng}$ treats all sustainability indicators equally, while the \textit{Prioritized Green Loss} $\mathcal{L}_{pg}$ assigns different priorities to each indicator.
The \textit{Non-prioritized Green Loss $\mathcal{L}_{ng}$} is then defined as:
\begin{equation}
    \mathcal{L}_{ng} = -\sum_{j=1}^n log(\sigma(  (g_j^{i_1}-g_j^{i_2})(\widetilde{y}_{i_1}- \widetilde{y}_{i_2}))).
\end{equation}

The \textit{Prioritized Green Loss $\mathcal{L}_{pg}$} is designed to recognize the varying significance of different sustainability indicators. For example, in food recommendations, factors like health and nutrition might be considered more crucial than the environmental impact of food production. Therefore, $\mathcal{L}_{pg}$ is formulated to incorporate these manually defined priorities.
Inspired by \cite{zhang2023targeted}, we propose considering an indicator only if all the indicator values with higher priorities exceed predefined thresholds. Specifically, assuming a priority order of $g_1>g_2>\cdots>g_n$, we first compare the values of $g_1$ for items $i_1$ and $i_2$. An indicator $g_j$ is then considered valid if, for all $g_k$ with $k < j$, the condition $min(g_k^{i_1}, g_k^{i_2}) \geq \beta_k$ is satisfied, where $\beta_k$ represents the threshold for the indicator $g_k$. $\mathcal{L}_{pg}$ is defined as:
\begin{equation}
    \mathcal{L}_{pg} = -\sum_{j=1}^n D(j)\cdot log(\sigma(  (g_j^{i_1}-g_j^{i_2})(\widetilde{y}_{i_1}- \widetilde{y}_{i_2}))).
    \label{equa:prioritized_green_loss}
\end{equation}
where
\begin{equation}
    D(j) = \left\{
             \begin{array}{ll}
                  1 &\text{if  } j=1 \text{  and } min(g_1^{i_1}, g_1^{i_2}) < \beta_1\\[7pt]
                  1 &\text{if  } j>1 \text{  and } 
                        \begin{array}{l}
                                min(g_j^{i_1}, g_j^{i_2}) < \beta_j; \\[2pt]
                               min(g_k^{i_1}, g_k^{i_2}) \geq \beta_k, \\[2pt]
                               \forall k\in \{1, ..., j-1\}
                        \end{array}  \\
             0 &\text{else}
             \end{array}
\right.
\end{equation}

\begin{table*}
    \centering
    \caption{Performances of different methods for Top-N recommendation. The best results are bold, and the second-best are underlined. A lower EIS indicates greater environmental friendliness, whereas higher NIS and HMI values denote more nutritious and healthier food, respectively.}
    \resizebox{2\columnwidth}{!}{
\begin{tabular}{ccccccccccccc} \toprule
Top-N & Metrics & BPR & KNN & SHT & STOSA & ICLRec & NOVA & CAFE & FDSA-CL & FHFRS & MSSR & \shortname{} ($\mathcal{L}_{np}$) \\ \midrule
\multirow{5}{*}{N=5} & HR & 0.0095 & 0.0107 & 0.0125 & 0.0113 & 0.0134 & 0.0137 & 0.0146 & \underline{0.0173} & 0.0165 & \textbf{0.0180} & 0.0159 \\
 & NDCG & 0.0080 & 0.0095 & 0.0098 & 0.0096 & 0.0125 & 0.0128 & 0.0137 & \underline{0.0157} & 0.0155 & \textbf{0.0164} & 0.0151 \\
 & EIS↓ & 89.58 & \textbf{83.70} & 96.43 & 113.92 & 97.78 & 89.67 & \underline{88.16} & 116.43 & 90.81 & 116.33 & 109.35 \\
 & NIS↑ & 32.59 & 31.23 & 33.83 & 35.51 & 35.26 & \underline{36.78} & \textbf{37.10} & 36.39 & 30.38 & 34.39 & 33.70 \\
 & HMI↑ & 44.56 & 44.94 & 40.56 & 35.38 & 35.31 & 44.69 & 43.50 & 43.01 & 44.81 & \textbf{45.99} & \underline{45.82}\\
 \midrule
\multirow{5}{*}{N=10} & HR & 0.0164 & 0.0188 & 0.0213 & 0.0207 & 0.0224 & 0.0243 & 0.0242 & 0.0269 & 0.0264 & \underline{0.0275} & \textbf{0.0285} \\
 & NDCG & 0.0112 & 0.0132 & 0.0149 & 0.0141 & 0.0168 & 0.0176 & 0.0181 & 0.0202 & 0.0201 & \underline{0.0208} & \textbf{0.0210} \\
 & EIS↓ & 88.82 & 84.90 & 86.32 & 88.56 & 104.54 & 88.38 & \textbf{79.43} & 92.43 & 87.11 & 83.79 & \underline{83.08} \\
 & NIS↑ & 31.79 & 31.15 & 32.16 & \underline{34.87} & 31.10 & \textbf{35.85} & 33.84 & 34.80 & 30.23 & 31.10 & 31.28 \\
 & HMI↑ & 42.10 & 42.91 & 43.98 & 43.88 & 43.46 & 43.36 & \underline{44.11} & 43.30 & 43.82 & 44.05 & \textbf{44.26}
 \\ \midrule
\multirow{5}{*}{N=20} & HR & 0.0269 & 0.0296 & 0.0344 & 0.0335 & 0.0343 & 0.0372 & 0.0399 & 0.0426 & 0.0390 & \underline{0.0437} & \textbf{0.0472} \\
 & NDCG & 0.0150 & 0.0171 & 0.0191 & 0.0187 & 0.0211 & 0.0222 & 0.0238 & 0.0259 & 0.0247 & \underline{0.0266} & \textbf{0.0279} \\
 & EIS↓ & 85.17 & 80.93 & 85.63 & 89.24 & 80.82 & 81.89 & \textbf{70.86} & 78.48 & 77.10 & 76.98 & \underline{76.42} \\
 & NIS↑ & 30.83 & 30.67 & 31.36 & 32.39 & 32.19 & \textbf{35.27} & 33.22 & 33.45 & 31.13 & 33.48 & \underline{33.78} \\
 & HMI↑ & 43.91 & 42.75 & 42.95 & 43.62 & 42.26 & 43.19 & 42.38 & 43.05 & 43.82 & \underline{44.36} & \textbf{44.67}\\ \bottomrule
\end{tabular} 
}
    \label{table:performances_all}
\end{table*}

\section{Experiment Settings}
\subsection{Dataset}
To evaluate the effectiveness of \shortname{}, we use a food recommendation dataset \cite{zhang2024greenrec} which contains 6290 users, 74324 recipes, and 316116 interactions. Each user or recipe has at least 10 interactions. Each recipe is labeled with three widely used sustainability indicators: 1) Environmental Impact Score (EIS) \cite{clark2022estimating, gephart2021environmental}, where a lower score indicates greater environmental friendliness; 
2) Nutritional Impact Score (NIS) \cite{clark2022estimating}, with a higher NIS score denoting more nutritious food; and
3) Healthy Meal Index (HMI) \cite{kasper2016healthy}, where a higher HMI score suggests a healthier meal. For dataset splitting, we employ a leave-one-out method.
\subsection{Baselines}
We select the following baselines:
\begin{itemize}
    \item \textbf{BPR} \cite{rendle2012bpr}, a traditional model that ranks items by maximizing the predicted score difference between a user’s preferred and non-preferred items.
    \cite{wang2006unifying}, a collaborative filtering method that recommends items by selecting items similar to user previously interacted items.
    \item \textbf{SHT}
    \cite{xia2022self}, a non-sequential baseline that enhances conventional matrix factorization with a hypergraph transformer network and generative self-supervised data augmentation .
    \item \textbf{STOSA}
    \cite{fan2022sequential}, which embeds each item as a stochastic Gaussian distribution, and forecasts the next item with a self-attention mechanism.
    \item \textbf{ICLRec}
    \cite{chen2022intent}, which learns users’ behaviors from unlabeled user historical interactions and is optimized through contrastive self-supervised learning.
    \item \textbf{NOVA}
    \cite{liu2021noninvasive}, which introduces a non-invasive attention mechanism to inject side-information into BERT structure for better attention distribution.
    \item \textbf{CAFE}
    \cite{li2022coarse}, which explicitly models user preference by fusing fine-grained item representations and coarse-grained side-information representations.
    \item \textbf{FDSA-CL}
    \cite{hao2023feature}, which introduces independent self-attention layers for item and feature representations and leverage contrastive method for training.
    \item \textbf{MSSR}
    \cite{lin2024multi}, which leverage self-attention machinism to capture item-feature and feature-feature correlation for both item and side-information prediction.
    \item \textbf{FHFRS} \cite{rostami2023towards}, which is a post-process green food recommendation model, and we adopt \textbf{MSSR} as its framework.
\end{itemize}

\subsection{Implementation Details}
In our experiment, we select learning rates from [0.0001, 0.01], embedding size $d$ from\{16, 32, 64\}, and batch size from \{32, 64, 128, 256\}. L2 regularization is applied when computing the loss function. 
For hyperparameters, we select $\alpha$ from $[0.5, 1]$ for the loss function. In prioritized green loss, we select hyperparameter $\beta_{EIS}$ from $[70, 120]$, $\beta_{NIS}$ from $[30, 50]$, and $\beta_{HMI}$ from $[30, 50]$. 
\footnote{Our code is available: https://github.com/JingXiaoyi/GRAPE.}

\subsection{Evaluation Metrics}
To evaluate the performances of Top-N sequential recommendation, we leverage the hit ratio (HR) and normalized discounted cumulative gain (NDCG) for $N=5,10,20$. We also calculate the average values of each sustainability indicator in the recommended list.


\begin{figure*}[ht]
    \centering
    \begin{minipage}{0.32\textwidth} 
        \subfloat[]{\includegraphics[width=\textwidth]{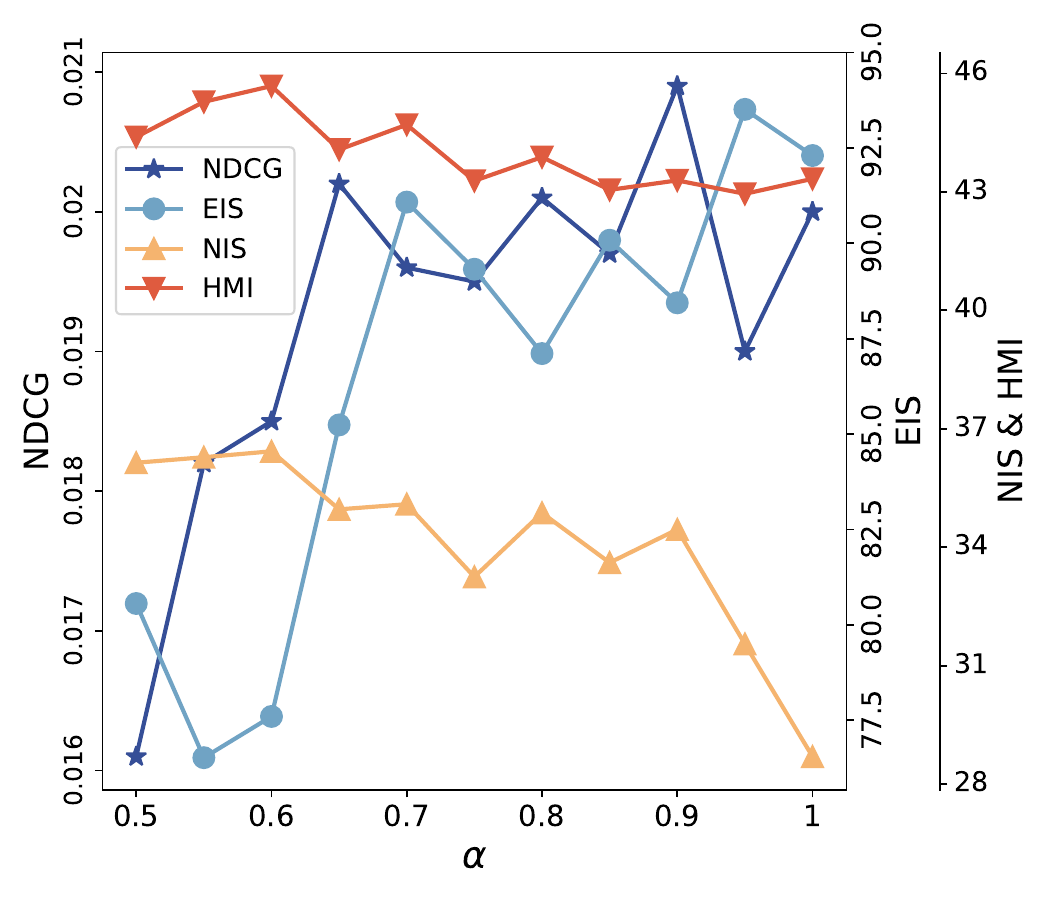}\label{fig:ablation_alpha}}
        \\
        \subfloat[]{\includegraphics[width=\textwidth]{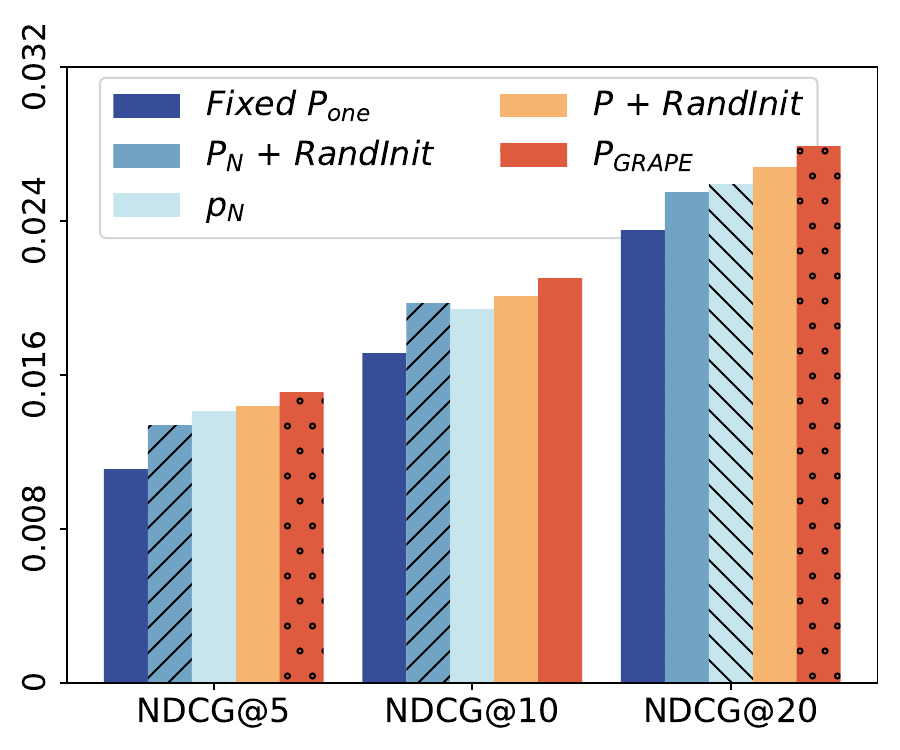}\label{fig:init}}
    \end{minipage}
    \begin{minipage}{0.65\textwidth} 
    \subfloat[Performance on NDCG@10]{\includegraphics[width=.49\columnwidth,trim=130 30 75 65,clip]{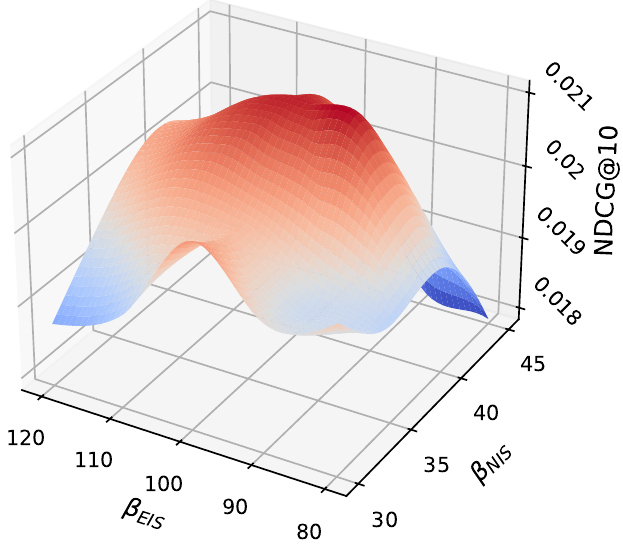}\label{fig:3d_ndcg}}
        \hfill
        \subfloat[Performance on EIS↓@10]{\includegraphics[width=.49\columnwidth,trim=130 30 75 65,clip]{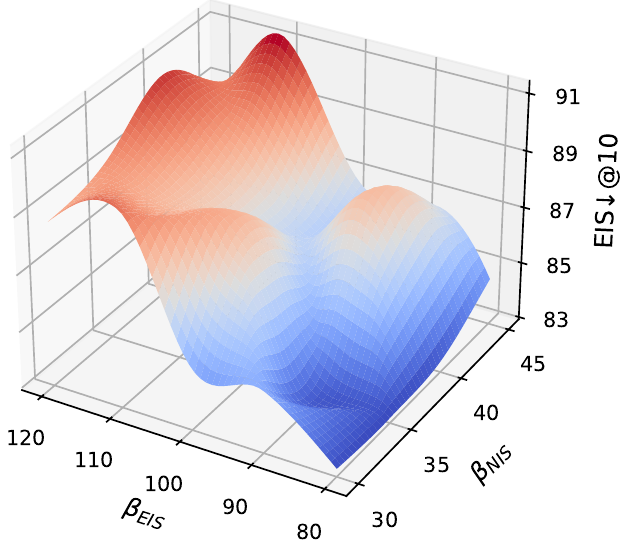}\label{fig:3d_env}}
        \\
        \subfloat[Performance on NIS↑@10]{\includegraphics[width=.49\columnwidth,trim=130 30 75 65,clip]{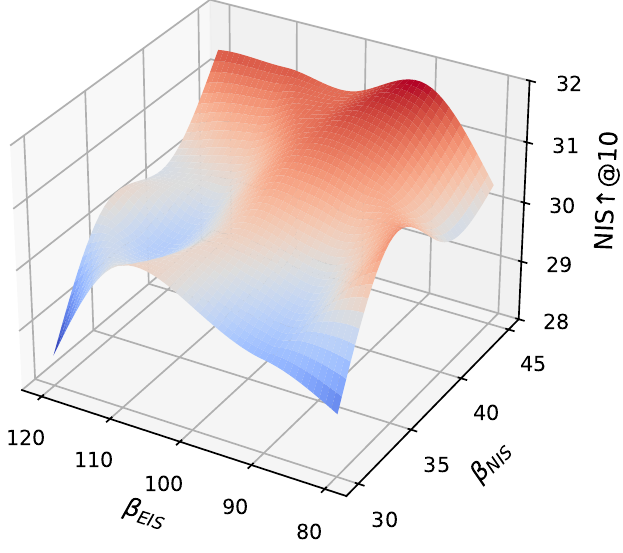}\label{fig:3d_nutri}}
        \hfill
        \subfloat[Performance on HMI↑@10]{\includegraphics[width=.49\columnwidth,trim=130 30 75 65,clip]{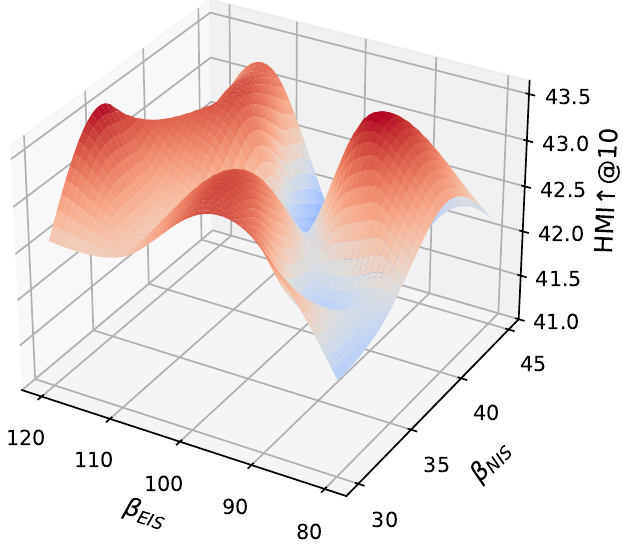}\label{fig:3d_meal}}
        \label{fig:3d_all}
    \end{minipage}
    
    \caption{Ablation studies: (a) Top-10 performances of \shortname{} with different hyperparameter $\alpha$ for the \textit{Non-prioritized Green Loss}. (b) Performances of \shortname{} with different user attention matrix. (c)-(f) Performances of recommendation accuracy and sustainability under varying $\beta_{EIS}$ and $\beta_{NIS}$ in the \textit{Prioritized Green Loss}, with the priority sequence set as EIS$>$NIS$>$HMI.}
    \label{fig:enter-label}
\end{figure*}

\section{Results and Analysis}
\subsection{Overall Performance}
Table \ref{table:performances_all} demonstrates the top-N recommendation performances of \shortname{} and all baselines. From the results, we make the following observations. First, in terms of recommendation accuracy, \shortname{} consistently attains the top position across all NDCG and most HR metrics when N=10 and N=20. For instance, \shortname{} leads the best baseline MSSR by 8.01\% in HR@20 and 4.89\% in NDCG@20. These findings demonstrate that \shortname{} can effectively model user preferences and suggest that integrating sustainability indicators may further improve recommendation accuracy.

Second, the food recommended by \shortname{} consistently achieves higher sustainability scores than those recommended by baseline models, especially against state-of-the-art models with notable top-N recommendation accuracy. This highlights \shortname{}'s ability to adaptively recommend sustainable items that align with user preferences. While traditional methods like KNN also recommend items with relatively high sustainability scores, they tend to achieve lower recommendation accuracy.


In short, our \shortname{} outperforms all state-of-the-art baselines in both recommendation accuracy and sustainability indicators. This demonstrates that recommending greener foods does not compromise the accuracy of recommendations and can, in fact, enhance them simultaneously.



\subsection{Effectiveness of Green Loss}
\label{Sec:Effect_of_NPGLoss}
To assess the effectiveness of our proposed Green Loss, we perform an ablation study using the Non-prioritized Green Loss on the hyperparameter $\alpha$ in Equation \ref{Equa: all_loss}. A higher $\alpha$ value indicates a lower proportion of green loss. The results are shown in Figure \ref{fig:ablation_alpha}.

Consistent with our assumption, as $\alpha$ increases, the recommendation list consistently displays lower values across all sustainability indicators. This suggests that our proposed green loss can effectively encourage the model to make greener recommendations.

We also observe that as $\alpha$ increases, the accuracy of recommendations first rises rapidly and then begins to fluctuate, with peak performance occurring at $\alpha=0.9$. These results highlight the need to balance the model's focus on user preferences and item sustainability. Although making greener recommendations contributes to higher recommendation accuracy, prioritizing users' personal preferences remains paramount to the recommendation process.


\subsection{Effectiveness of User Attention Matrix}
In this work, we assume that users have varying preferences and acceptance levels for sustainable items. To address this, we design a user attention matrix $P$ to explicitly capture these preferences. We initialize $P$ based on insights derived from users' historical interactions. To validate our assumption and assess the effectiveness of our proposed initialization methods, we implement several variants of $P$ using different initialization approaches:
\begin{itemize}
    \item \textit{Fixed $P_{one}$}: We replace $P$ with frozen all-ones matrix $P_{one}=\bm{1}^{v\times (n+1)}$, indicating that there are no distinctions in personal preferences among items or sustainability indicators.
    \item \textit{$P_N$ + RandInit}: We apply a consistent vector $\bm{p} \in \mathbb{R}^{(n+1)}$ to populate each row of the matrix $P_{N}\in \mathbb{R}^{v \times(n+1)}$. $\bm{p}$ is randomly initialized. This variant indicates that all users share a uniform preference towards sustainability.

    \item \textit{$P_N$}: Similar to \textit{$P_N$ + RandInit}, we apply a consistent vector $\bm{p} \in \mathbb{R}^{(n+1)}$ to represent all users's preferences. We then initialize $\bm{p}$ using the average and standard deviation of the sustainability values from items interacted with by all users.
    \item \textit{$P$ + RandInit}: We use randomly initialization for the user attention matrix $P$.
    \item \textbf{$P_\shortname{}$}: This is the default setting for \shortname{}.
\end{itemize}

The results are illustrated in Figure \ref{fig:init}. First, we observe that Fixed $P_{one}$ performs significantly worse than all other variations, supporting our assumption that users have different preferences among items or sustainability indicators. Moreover, when comparing the non-personalized variations, \textit{$P_N$ + RandInit} and \textit{$P_N$}, with the personalized ones, \textit{$P$ + RandInit} and \textbf{$P_\shortname{}$}, we find personalized group performs better. This suggests that users' preferences for different attributes vary from person to person. Furthermore, \textbf{$P_\shortname{}$} outperforms all other variations, demonstrating the effectiveness of initialization using the average and standard deviation of sustainability values from users' interacted items.


\begin{table}
    \centering
    \caption{Top-10 performance of \shortname{} with \textit{Non-prioritized Green Loss} ($\mathcal{L}_{ng}$), and \textit{Prioritized Green Loss} ($\mathcal{L}_{pg}$) applying different priority orders.
    $\CIRCLE$ denotes the highest priority, $\LEFTcircle$ denotes the second priority, and $\Circle$ denotes the lowest priority. The best results for each evaluation metric are bold.}
    \resizebox{\columnwidth}{!}{
\begin{tabular}{c|ccccc} \toprule
Model & HR  & NDCG& EIS↓   & NIS↑   & HMI↑   \\ \midrule
\multirow{6}{*}{\shortname{} ($\mathcal{L}_{pg}$)} 
& 0.0285  & \textbf{0.0210}  & \CIRCLE 83.08  & \LEFTcircle 31.28  & \Circle 44.26  \\
& \textbf{0.0288}  & 0.0194  & \CIRCLE \textbf{79.47} & \Circle 30.38  & \LEFTcircle 42.57  \\
& 0.0276  & 0.0205  & \LEFTcircle 80.20 & \CIRCLE 31.85 & \Circle 41.44  \\
& 0.0264  & 0.0198  & \Circle 86.31 & \CIRCLE \textbf{33.40} & \LEFTcircle 42.41  \\
& 0.0286  & 0.0201  &\LEFTcircle 89.67 & \Circle 33.02  & \CIRCLE 41.68  \\
& 0.0285  & 0.0202 & \Circle 79.99  & \LEFTcircle 32.56  & \CIRCLE \textbf{44.27} \\
\midrule
\shortname{} ($\mathcal{L}_{ng}$)  & 0.0285 & 0.0206  & 81.41  & 30.92  & 43.19 \\ \bottomrule
\end{tabular}
}
    
    \label{table:performances_prioritized_loss}
\end{table}

\subsection {Ablation Study for Prioritized Green Loss}
Our prioritized green loss $\mathcal{L}_{pg}$ allows \shortname{} to make greener recommendations by prioritizing sustainability indicators in a specific order. To evaluate the effectiveness of $\mathcal{L}_{pg}$, we experiment with various priority orders and set different thresholds $\beta$ for the indicators within $\mathcal{L}_{pg}$. 

Table \ref{table:performances_prioritized_loss} displays the performances when using different priority orders for three sustainability indicators. In line with our expectations, the sustainability indicator assigned the highest priority consistently achieves the best performance within its category. 
Additionally, we observe that setting either NIS or HMI as the highest priority results in relatively high performance for the other indicator as well. This may be due to a high correlation between the food nutritional level and the healthy meal index.



Then we select the priority EIS$>$NIS$>$HMI and conduct ablation study on the sustainability thresholds $\beta_{EIS}$ and $\beta_{NIS}$. It is important to note that a smaller $\beta_{EIS}$ or a larger $\beta_{NIS}$ imposes stricter constraints on the corresponding sustainability indicator, which is expected to yield more sustainable recommendations within that category. The results are shown in Figure \ref{fig:3d_ndcg}-\ref{fig:3d_meal}, with each graph representing a different evaluation metric.

As shown in Figure \ref{fig:3d_env} and \ref{fig:3d_nutri}, we observed that when the constraints on a specific indicator are tightened, the performances of the corresponding indicator improves generally. Such findings demonstrate that our loss function effectively adjusts the model's focus across different indicators.

From Figure \ref{fig:3d_ndcg}, we observe that the recommendation accuracy first increases and then decreases with the rise in either $\beta_{EIS}$ or $\beta_{NIS}$. This may be attributed to the model's tendency to focus primarily on a particular emphasized sustainability indicator, which can sometimes come at the expense of overall recommendation performance. It underscores the necessity to strike a balance between user preferences and item sustainability to optimize recommendation effectiveness.


\section{Conclusion}

In conclusion, recommendation systems have significant potential to encourage more sustainable choices among users. However, most existing methods focus solely on recommendation accuracy, often overlooking the potential impact of recommending more sustainable items. To address this oversight, we introduce the Green Food Recommendation task and propose a novel method called \longname{} (\shortname{}). \shortname{} not only models users' evolving preferences for items but also their willingness to choose sustainable foods. Extensive experiments demonstrate the superiority of \shortname{}, showing that it successfully balances recommendation accuracy with enhanced sustainability attributes of the recommended foods. We believe that our research lays a strong foundation to encourage future studies in this field.

\section{Acknowledgments}

This research is supported by the RIE2025 Industry Alignment Fund – Industry Collaboration Projects (IAF-ICP) (Award I2301E0026), administered by A*STAR, as well as supported by Alibaba Group and NTU Singapore through Alibaba-NTU Global e-Sustainability CorpLab (ANGEL). It is also supported by the Joint NTU-UBC Research Centre of Excellence in Active Living for the Elderly (LILY) and College of Computing and Data Science (CCDS) at NTU Singapore.

\bibliography{aaai25}

\appendix
\newpage
1
\newpage
\subsection{Implementation Details}
In our experiment, we select learning rates from [0.0001, 0.01], embedding size $d$ from\{16, 32, 64\}, and batch size from \{32, 64, 128, 256\}. L2 regularization is applied when computing the loss function. 
For hyperparameters, we select $\alpha$ from $[0.5, 1]$ for the loss function. In prioritized green loss, we select hyperparameter $\beta$ for EIS $\beta_{EIS}$ from $[70, 120]$, for NIS $\beta_{NIS}$ from $[30, 50]$, and for HMI $\beta_{HMI}$ from $[30, 50]$. 
Our code and data have been made publicly available\footnote{https://github.com/AAAI2025Author/GRAPE}.

\subsection{Example for \textit{Prioritized Green Loss}}
To illustrate the operation principle of \textit{Prioritized Green Loss} more clearly, we have provided several examples in Table \ref{table:Appendix_Example}. 
In Case 1, the EIS indicator with the highest priority is considered valid because $g^{i_2}_{EIS}=86$ is not greener than threshold $\beta_{EIS}=80$. A similar scenario occurs in Case 3 where the sustainability indicator HMI with the highest priority is valid. In Case 2, both EIS values with the highest priority are greener than threshold $\beta_{EIS}=90$. Afterward, $g^{i_2}_{NIS}=34$ with the second priority is not greener than $\beta_{NIS}=35$, making the indicator NIS valid.

\begin{table}[ht]
    \centering
    \caption{Example for prioritized green loss. $D(\cdot)=1$ denotes the corresponding sustainability indicator is considered valid in the given case, while $D(\cdot)=0$ denotes the indicator is not considered valid.}
    \label{table:Appendix_Example}
    \resizebox{\columnwidth}{!}{
   \begin{tabular}{c|ccc} \toprule   
Indicator                    & EIS↓ & NIS↑ & HMI↑ \\ \midrule
Item1                        & 74   & 37   & 42   \\
Item2                        & 86   & 34   & 46   \\ \midrule \midrule
\multicolumn{4}{c}{Case 1}                        \\ \midrule
Priority Order & $1^{st}$  & $2^{nd}$  & $3^{rd}$  \\
Threshold & $\beta_{EIS}=80$   & $\beta_{NIS}=30$   & $\beta_{HMI}=45$   \\
$D(\cdot)$                          & 1    & 0    & 0    \\ \midrule \midrule
\multicolumn{4}{c}{Case 2}                        \\ \midrule
Priority Order & $1^{st}$  & $2^{nd}$  & $3^{rd}$  \\
Threshold & $\beta_{EIS}=90$   & $\beta_{NIS}=35$   & $\beta_{HMI}=45$   \\
$D(\cdot)$                          & 0    & 1    & 0    \\ \midrule \midrule
\multicolumn{4}{c}{Case 3}                        \\ \midrule
Priority Order & $3^{rd}$  & $2^{nd}$  & $1^{st}$  \\
Threshold & $\beta_{EIS}=80$   & $\beta_{NIS}=30$   & $\beta_{HMI}=45$   \\
$D(\cdot)$                          & 0    & 0    & 1   \\ \bottomrule
\end{tabular}
}
\end{table}

\subsection{Dataset Details}
\begin{figure}[ht]
	\centering
	\subfloat[Statistics of EIS↓ for user's interacted items.] {\includegraphics[width=.8\columnwidth, trim=12 7 12 11, clip]{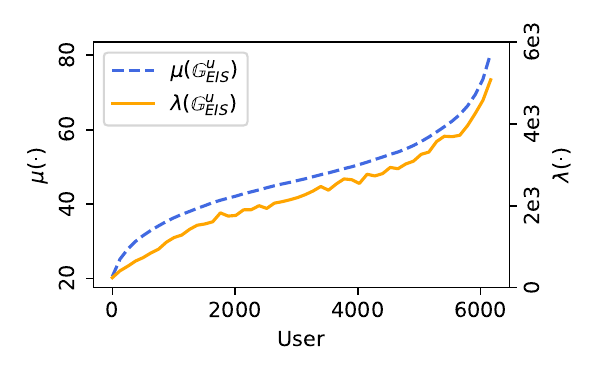} \label{fig:fig_boxplot_tag_in_all_dataset_top_10} \label{fig:Appendix_Statistic_Dataset_EIS}}
    \\
	\subfloat[Statistics of NIS↑ for user's interacted items.] {\includegraphics[width=.8\columnwidth, trim=12 7 12 11, clip]{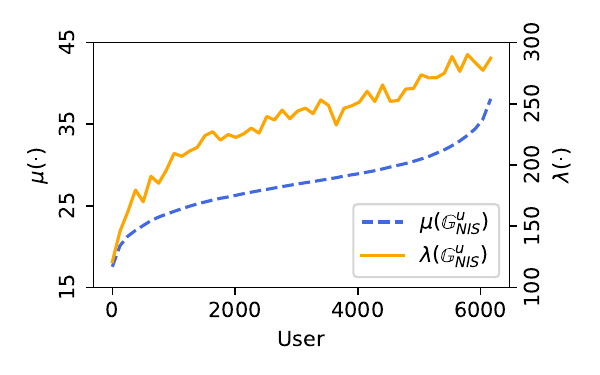}
    \label{fig:fig_boxplot_tag} \label{fig:Appendix_Statistic_Dataset_NIS}}
    \\
    \subfloat[Statistics of HMI↑ for user's interacted items.]{\includegraphics[width=.8\columnwidth, trim=12 7 12 10, clip]{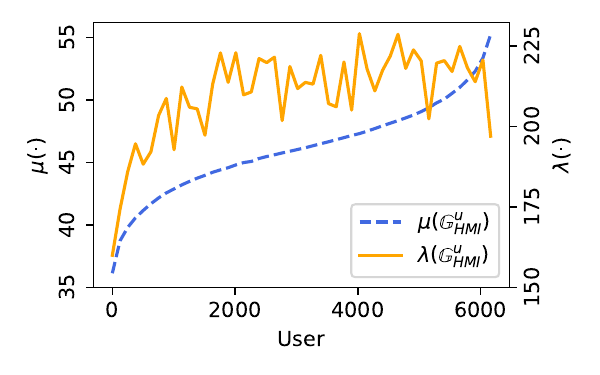} \label{fig:Appendix_Statistic_Dataset_HMI}}

	\caption{Statistic for mean ($\mu(\cdot)$) and variance ($\lambda(\cdot)$) of sustainability indicator sequence $\mathbb{G}^u_{EIS}$, $\mathbb{G}^u_{NIS}$, $\mathbb{G}^u_{HMI}$ for user's interacted items $\mathbb{I}^u$. The results are displayed in ascending order based on the mean values of the corresponding sustainability indicators for each user.
    }
	\label{fig:Appendix_Statistic_Dataset}
\end{figure}

Given user $u$'s interacted item list $\mathbb{I}^u$ and the corresponding sustainability indicator list $\mathbb{G}^u_{EIS}$, $\mathbb{G}^u_{NIS}$, and $\mathbb{G}^u_{HMI}$, we calculate the mean ($\mu(\cdot)$) and variance ($\lambda(\cdot)$) of each sequence, respectively. The results are shown in Figure \ref{fig:Appendix_Statistic_Dataset}. The x-axis represents the users. We make the following observation. 
First, we observe significant variation in the average sustainability indicators of items interacted with by different users, which further indicates that users' green preferences are highly diverse.
Second, as shown in Figure \ref{fig:Appendix_Statistic_Dataset_NIS} and Figure \ref{fig:Appendix_Statistic_Dataset_HMI}, as users shift towards healthier eating habits, the variance in the health index of their recipes first increases, followed by fluctuations, and potentially a downward trend in the HMI index. This trend may be due to certain users prioritizing health and nutritional balance, which results in more consistent eating habits over time. Third, as shown in Figure \ref{fig:Appendix_Statistic_Dataset_EIS}, users consuming more environmentally friendly diets exhibit relatively low variance in terms of EIS, which likely indicates vegetarian eating patterns.

\end{document}